\documentclass[preprint,12pt]{elsarticle}

\usepackage{amssymb}
\usepackage{subcaption}
\usepackage{xurl}
\usepackage{xcolor}

\journal{Nuclear Inst. and Methods in Physics Research, A }

\begin{document}
\begin{frontmatter}



\title{Modelling the response of a CsI(Tl)-PiN photodiode Microscintillator Detector}


\author[1]{Justin Tabbett\corref{cor}}
\ead{jt16596@bristol.ac.uk}
\author[1]{Karen L. Aplin}
\affiliation[1]{organization={Faculty of Engineering, University of Bristol},
            addressline={Queen's Building, University Walk}, 
            city={Bristol},
            postcode={BS8 1TR}, 
            country={United Kingdom}}
\cortext[cor]{Corresponding author}

\begin{abstract}
The full instrument response of a superminiaturised CsI(Tl)-PiN photodiode radioactivity detector, intended for deployment on a meteorological radiosonde, has been modelled by combining a physics-based model of the sensor  with the detector circuit response, obtained via an LTspice simulation. The model uses the incident energy of a gamma ray as an input, and produces the pulse expected from the detector. The detector response was verified by comparing the simulated energy calibration with a laboratory {source}. The measurement circuit is found to control the minimum detectable energy of 26 keV, and the maximum detectable energy is $\sim$10 MeV. The energy sensitivity of the PiN detector is {0.29 $\pm$ 0.02} mV/keV in the 0-800 keV range. The simulation and laboratory calibrations were consistent to better than 5\% over the {calibration} range of the instrument.
\end{abstract}




\end{frontmatter}


\section{Introduction}
\label{section:Introduction}
There is a lack of readily available instrumentation to study ionisation in the atmosphere, and the effects of energetic particles on weather and climate 
\cite{Mironova2015}. The creation of atmospheric ions by galactic cosmic rays, solar UV radiation, and electron precipitation events means that the effects of these ions manifest in different regions of the atmosphere. For instance, neutron monitoring stations provide a global understanding of galactic cosmic ray intensities, however at altitudes below 5 km, it has been shown that there is little correlation between neutron counts and measured ionisation rates \cite{HARRISON2014203}. Predominantly, satellites have been used for primary particle detection \cite{BAUMGARDNER201110}, and ground-based instruments for secondary particle detection, however the intermediary region creates an opportunity for a new, balloon-borne detector. 

A novel microscintillator ionisation detector, here called the PiN detector, capable of measuring energy and count rate, has been developed for deployment on meteorological radiosondes (weather balloons). Hundreds of these balloons are launched daily for weather forecasting purposes, but as they are not routinely retrieved, and have limited capability to carry additional payloads, the cost is limited to a few hundred pounds and the mass to tens of grams. Geigersondes are suitable for balloon applications but do not offer energy detection \cite{HARRISON2014203}. A miniaturised CsI(Tl) scintillator coupled to a PiN diode both meets the radiosonde power and mass requirements, and adds energy detection capability.
The detector was first deployed on a meteorological radiosonde in 2016 to investigate the transition region where surface-borne energetic particles cease to be the dominant source of ionising radiation, in favour of high-energy particles in the free troposphere \cite{KA2017}. During balloon deployment in 2018, the detector unexpectedly observed stratospheric X-rays, which was corroborated by NOAA POES spacecraft data \cite{KA2021}. 
Deployment of the detector is well-established, however the work presented in this paper analyses and discusses the instrumentation in more detail than previously. This will allow both for retrospective analysis of previous flight data and for future development.

\begin{figure}
    \centering
    \includegraphics[width=0.8\textwidth]{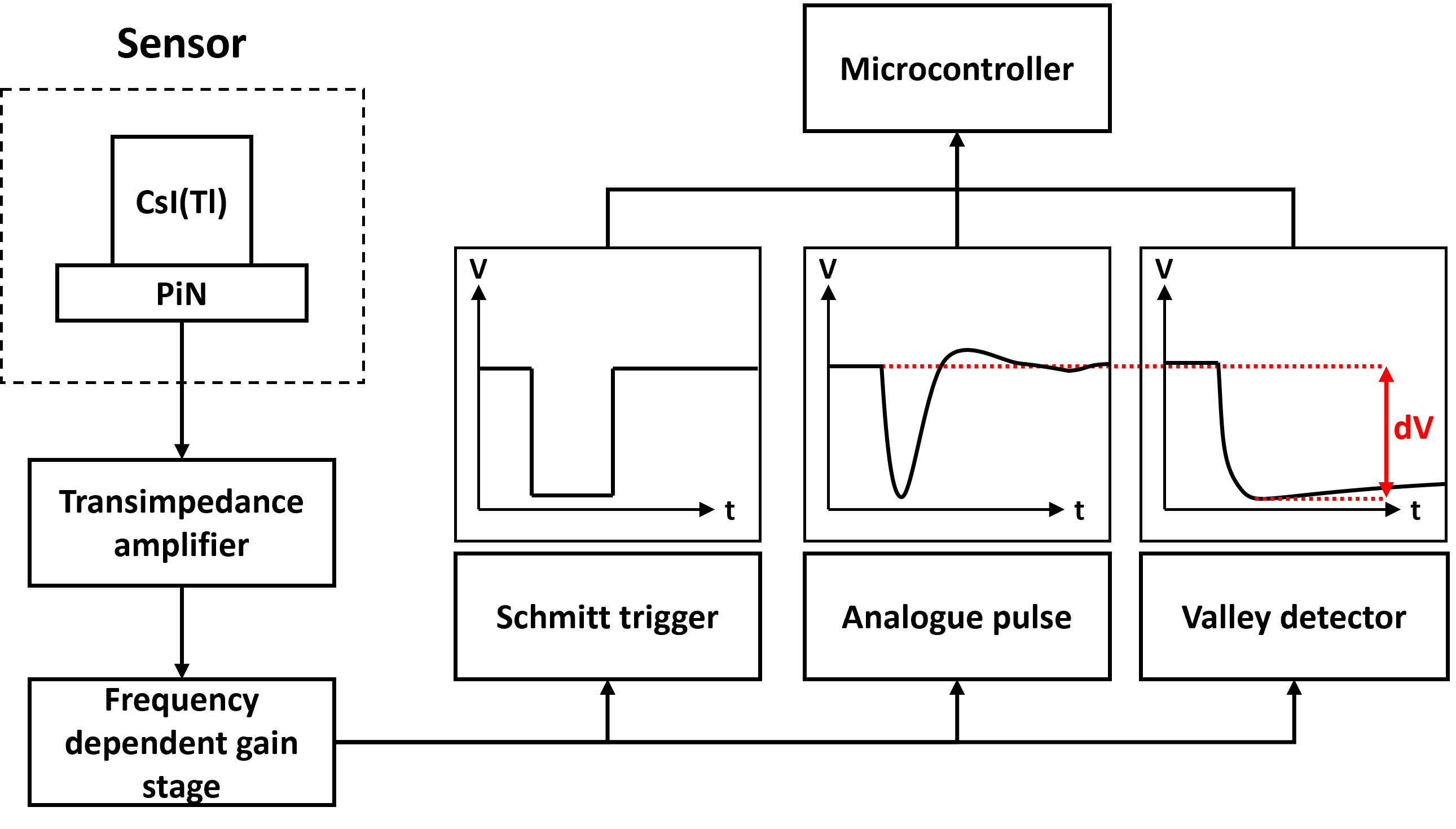}
    \caption{Block diagram showing the detection process and illustration of the pulse height. The voltages for the pulse height are measured in the microcontroller.}
    \label{fig:blockdiag}
\end{figure}

The PiN detector sensor is an Advatech CsI(Tl) scintillator, measuring 10$\times$10$\times$8 mm\textsuperscript{3}, coupled to a silicon PiN RD100 photodiode \cite{adva,osipin}. There are two stages in particle detection within the sensor. First, incident radiation generates a light pulse in the scintillator, described by characteristic decay times. The second stage of particle detection involves the conversion of a light pulse to a current pulse. The magnitude of the current pulse is proportional to the energy of the incident ionising radiation. The next stage of particle detection involves the current passing through the electronics of the detector. Figure \ref{fig:blockdiag} shows the detection process where the current flows from the PiN photodiode to a transimpedance amplifier, located physically close to the photodiode on the board to minimise losses. The signal then passes through a frequency dependent gain stage, followed by the signal conditioning circuitry.

There are three key components in the signal conditioning circuit: a Schmitt trigger, an analogue pulse, and a negative peak detection circuit (valley detector). The trigger activates an interrupt routine in the PIC16F676 microcontroller code, signalling the microcontroller to measure the voltages on the analogue pulse and the valley signal. 

A pulse height, $dV$, is calculated by the difference between {a reference point on }the analogue pulse and the valley trace. The measured analogue pulse value is larger than the measured valley value, as the analogue pulse is a negative pulse.

Regarding the detector data collection, the microcontroller has a 10-bit analogue-digital converter (ADC), which with a 5 V supply, gives 5 V as 1023 ADC counts {(referred to in the following text as `ADC')}. The reference voltage level rests at $\sim$ 3.211 V (657 ADC). The pulse height should then not exceed 657 ADC. The microcontroller samples the reference and valley voltage values separately, timestamps them, and sends them to a serial output. 
 
We present, for the first time, a full-stack model of the detector response, from the initial interaction of the scintillator with ionising radiation, through to the production of signals which are passed to the microcontroller for measurement.  
A physics-based model written in Python emulates the response of the sensor, the signals from which are used as input to a Simulation Program with Integrated Circuit Emphasis (SPICE) simulation to emulate the electronics response of the detector. {The model therefore takes a single input energy and produces the corresponding pulse from the detector electronics - modelling the monoenergetic response}.

Software packages such as Geant4, FLUKA, and MCNPX are often used to simulate the interactions between ionising radiation and scintillator crystals \cite{FLUKA}. FLUKA was not suitable as the desired light pulse output needed to be in the time-domain, to combine with the photodiode response; whereas FLUKA produces quantities such as spatial energy depositions, emission spectra, and particle fluence, among others. Using Geant4 in combination with SPICE programs has been suggested as a method of investigating the overall response of a detector and would likely be a valid approach \cite{convolution}. However, as the scintillator is a commercial product with its typical response provided by the manufacturer, it would be unnecessary to model its response from first principles. 
Additionally, as the goal is simulation of the total response of the detector, the response of electronic components used in the signal conditioning circuit are as important as the detailed physics of the sensor itself. This motivates an analytical approach to the model, rather than focusing on the small differences likely to arise between a numerical and analytical approach \cite{POZZI2007629}.

\begin{figure}
    \centering
    \includegraphics[width=0.8\textwidth]{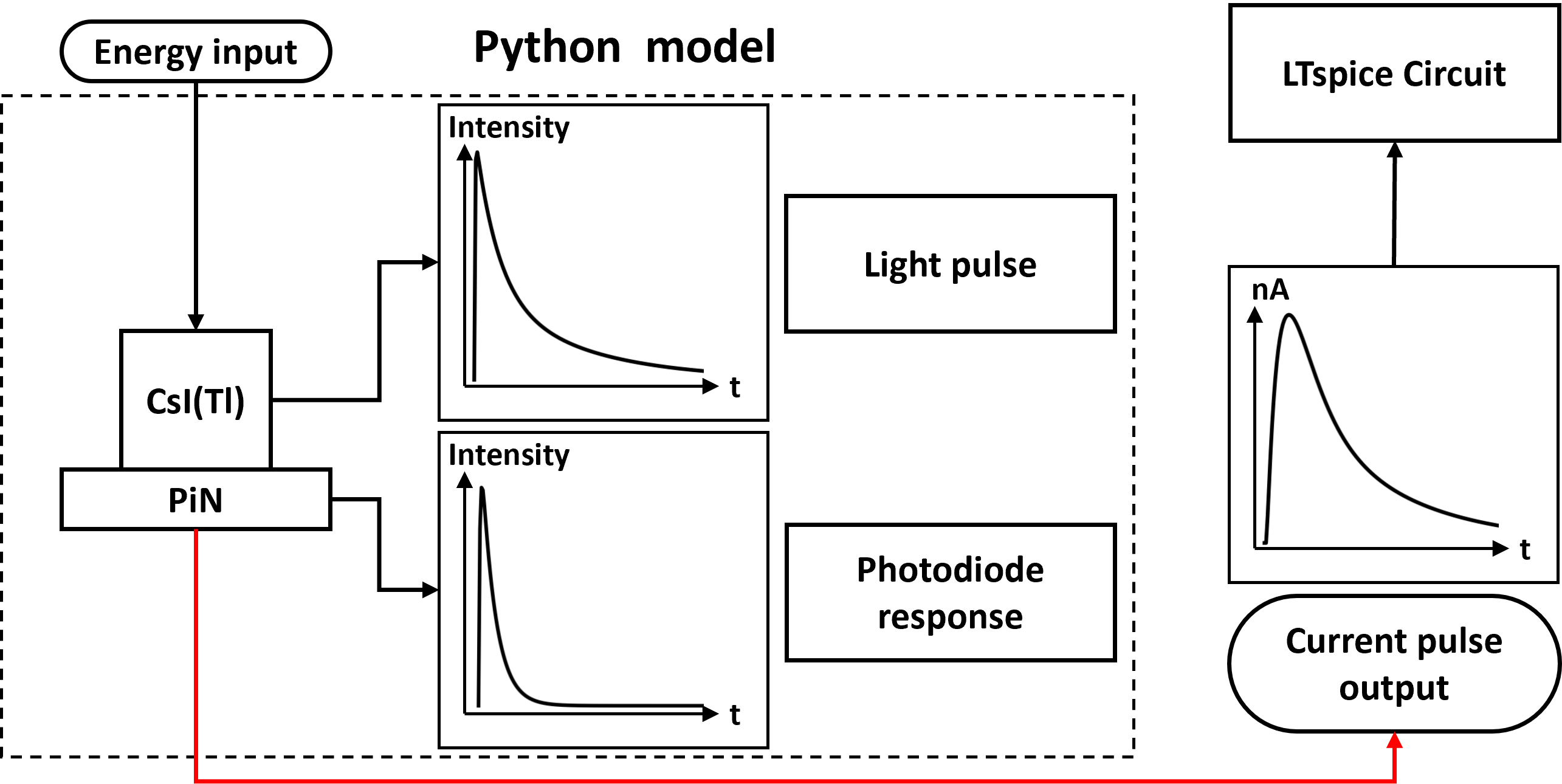}
    \caption{Model block diagram illustrating the pulse shapes obtained at each stage of the Python model. The light pulse has a shorter decay time than the PiN photodiode response. The two responses are combined using convolution to determine the shape of the current pulse output; subsequently used as the input for the LTspice simulation.}
    \label{fig:pyblock}
\end{figure}

Like the scintillator, the photodiode is governed by characteristic decay times (rise- and fall-times); however, as a commercial component, only typical rise times, at a specific reverse bias and for a given wavelength, are provided. Therefore, the rise time has been determined by using estimated parameters, and the fall time was determined during a tuning phase of the model development. Convolution has been used as the method of interfacing the scintillator and photodiode responses \cite{convolution}, ultimately producing a current pulse. Figure \ref{fig:pyblock} illustrates the detection process, noting the output from each stage of the Python model.
Finally, the SPICE simulation has been created using LTspice; the current pulse from the physics model is used as an input for the simulation. Relevant circuit components have been recreated in LTspice, allowing for analogue voltage traces to be measured.  

Accompanying the model and simulation, a laboratory calibration with a radioactive source is presented, serving as the primary method of validation for the model.
Section \ref{pymod} details the model method, with the scintillator response in Section \ref{scre}, the photodiode response in Section \ref{phre}, and the current pulse in Section \ref{cre}. Simulation of the detector electronics is detailed in Section \ref{ltsec}, and the laboratory calibration is explored in Section \ref{labcal}. Finally, a comparison of the model and detector is given in Section \ref{comp}.

\section{Model Method}
Modelling the PiN detector response was split into two sections: the physics model, and the electronics simulation. Within the physics model, the responses of the scintillator and photodiode are considered, resulting in their combination which produces a current pulse proportional to the incident radiation. The current pulse serves as an input for the electronics simulation. The simulation models the remainder of the detector until the data acquisition stage.
\label{pymod}
\subsection{CsI(Tl) Scintillator Response}
\label{scre}
To model the production of photons by the interaction of radiation with the scintillator, the following assumptions were made \cite{Valentine,958740}:
\begin{itemize}
    \item Total number of photons generated are determined by the light yield {- the number of photons emitted per unit energy ($\gamma\mathrm{/MeV}$) -} of the scintillator
    \item Photons are emitted over a range of wavelengths in different proportions, described by the emission spectrum
    \item The light pulse generated is characterised by at least two decay times
\end{itemize}
The scintillator emission decay curve is described by Equation \ref{scintshape}
\begin{equation}
\label{scintshape}
    L(t)=(1-\exp{[-t/\tau_{rs}])}-a_1(1-\exp{[-t/\tau_{1}])}-a_2(1-\exp{[-t/\tau_{2}])}
\end{equation}
where $\tau_{rs}$ is scintillator rise time, $\tau_{1,2}$ are the fast and slow decay times, respectively, and $a_{1,2}$ their proportions. {If additional decay components are desired, they can be added to Equation \ref{scintshape}, with the limit that the summation of proportions is equal to one.} In the instance of the CsI(Tl) crystal present in the PiN detector, the manufacturer reports a single decay time, therefore in the model $a_2=0$. The scintillator parameters are given in Table \ref{tab:params}, and {different numbers of decay components are considered in Section \ref{modcon}.} 

The model uses discretised wavelength steps to determine the energy generated by scintillation photons during an interaction. For each wavelength, $\lambda$, a number of photons, $N(\lambda)$ are produced described by the light yield $LY$, incident energy $E_{in}$, and emission spectrum proportion $S(\lambda)$:

\begin{equation}
    N(\lambda)=LY \cdot E_{in} \cdot S(\lambda)
\end{equation}

\begin{figure}
    \centering
    \includegraphics[width=0.75\textwidth]{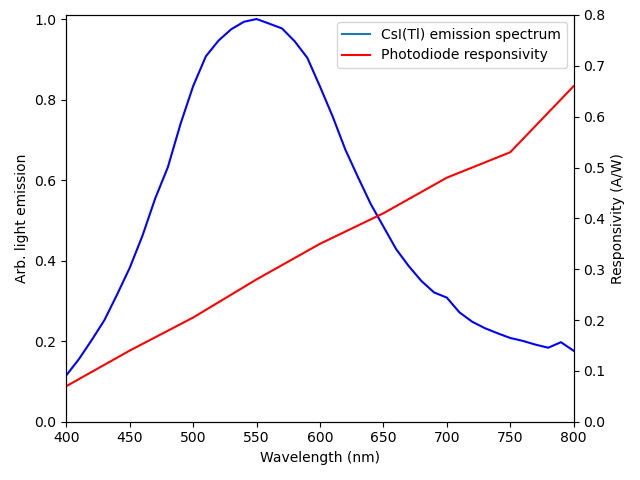}
    \caption{Digitised scintillator emission spectrum and photodiode responsivity.}
    \label{fig:emissrespon}
\end{figure}

The emission spectrum, shown in Figure \ref{fig:emissrespon}, was digitised using a Python-based plot digitiser and linear interpolation \cite{BELOGUROV2000254,Digtiser}. This processes allows for the scintillator emission spectrum and the photodiode responsivity to be evaluated at the same steps (every 10 nm) over the model's wavelength range (400-800 nm). 
Therefore, a summation of the energy of a single wavelength photon, $E_s= hc/\lambda$, multiplied by the corresponding number of photons, over this range, yields the total energy emitted by the scintillator for a single event.

\begin{equation}
    E_{Tot}=\sum_{\lambda=400}^{800} E_s(\lambda)\cdot N(\lambda)
\end{equation}

It is necessary to convert the energy emitted from the scintillator into a power quantity, because the photodiode responsivity relates incident optical power to an output photocurrent - discussed in Section \ref{cre}.
Therefore, {an energy dependent function (cutoff time)}, $t_{co}$, has been empirically determined. 
{
\begin{equation}
     t_{co}(E_{in})=(-4.0\pm0.2)\exp{[-E_{in}/(460\pm60)]}+(4.1\pm0.2)
\end{equation}
The cutoff time function returns a single value in $\mu$s. The function was determined by evaluating test values until a small ($\lesssim$2 \%) percentage difference was present between the model output and the expected output (determined by the laboratory calibration).}

The final expression for the power output from the scintillator is then 
{
\begin{equation}
    P(t)=\frac{E_{Tot}}{t_{co}(E_{in})}\cdot L(t)
\end{equation}
}
\begin{table}
    \caption{Scintillator Model Parameters. `N/A' means `not applicable'}
    \centering
    \begin{tabular}{|l|c|c|c|}
    \hline
        Parameter & Symbol & Value & Source \\
        \hline 
        Scintillator rise time & $\tau_{rs}$ & 50 ns & Estimated \cite{Sun_2011} \\
        Fast decay time & $\tau_1$ & 900 ns & Datasheet \cite{adva} \\
        Slow decay time & $\tau_2$ & N/A & N/A\\
        Cutoff time & $t_{co}$& - Variable & Empirical function\\
        Fast decay proportion & $a_1$ & 1 & Datasheet \cite{adva} \\
        Slow decay proportion & $a_2$ & 0 & N/A \\
        Light yield & LY & 54000 $\gamma$/MeV & Datasheet \cite{adva} \\
        \hline
    \end{tabular}
    \label{tab:params}
\end{table}

\subsection{PiN Photodiode Response}
\label{phre}
The temporal response of the photodiode is governed by the rise time $\tau_{rp}$, typically composed of: drift and diffusion times, and the RC time constant of the diode circuit. The relationship between the three components is given by:

\begin{equation}
    \label{photorise}
    \tau_{rp}=\sqrt{\tau_{RC}^2+\tau_{drift}^2+\tau_{diff}^2}
\end{equation}
where $\tau_{RC}$, $\tau_{drift}$, and $\tau_{diff}$ are the RC, drift times, and diffusion times, given in Equation \ref{eqn:RC}, \ref{eqn:drift}, and \ref{eqn:diff} respectively.

The derivation of the RC time constant and its dependencies are given in Equations \ref{eqn:RC}-\ref{eqn:rs} which denote the RC time constant, the junction capacitance $C_j$, the depletion width $W_d$, and the series resistance $R_s$ \cite{OSIappnote}:

\begin{equation}
\label{eqn:RC}
    \tau_{RC}=2.2 C_j (R_s+R_L)
\end{equation}
where $R_L=50 \Omega$ is the load resistance which has been estimated;

\begin{equation}
\label{eqn:cj}
    C_j=\frac{\epsilon_{Si}\epsilon_{0}A}{W_d}
\end{equation}
where $\epsilon_{Si} = 11.9$ is the relative permittivity of silicon, $\epsilon_{0}$ is the permittivity of free space, and $A= 100$ mm\textsuperscript{2} is the active area of the photodiode;
\begin{equation}
    \label{eqn:Wd}
    W_d=\sqrt{\frac{2\epsilon_{Si}\epsilon_{0}}{qN_n}(V_A+V_{Bi})}
\end{equation}
where $q$ is the charge of an electron, $V_A=12$ V is the applied bias, $V_{Bi}=0.65$ V is estimated value of the built-in bias of silicon {(which has typical forward bias between 0.6-0.7 V)}, and $N_n$ is the doping concentration;

\begin{equation}
    \label{eqn:rs}
    R_s=R_c+\frac{(W_s-W_d)\rho}{A}
\end{equation}
where {$\rho=(qN_n\mu)^{-1}$ is the resistivity of silicon} and $\mu$ is the electron mobility, $R_c$=0 $\Omega$ is the assumed contact resistance in the diode, and $W_s$ is the silicon substrate width.

The substrate width and doping concentration $N_n$ were estimated by considering the case where $V_A=100$ V, and the photodiode is completely depleted, meaning $W_s=W_d$. Using the previous assumption, Equation \ref{eqn:cj} and the typical value of the junction capacitance, 50 pF, under such conditions, the substrate width is estimated to be $W_s=210.7252$ $\mu$m. Following, an estimate for the doping concentration can be obtained using this result and Equation \ref{eqn:Wd}, giving $N_n=2.985 \times10^{18}$.

Additionally, the electron mobility of silicon $\mu$=1350 cm\textsuperscript{2}(Vs)\textsuperscript{-1} \cite{Knoll}.

The drift time \cite{OSIappnote} component is given by:
\begin{equation}
\label{eqn:drift}
    \tau_{drift}=\frac{W_d^2}{2\mu(V_A+V_{Bi})}
\end{equation}

The diffusion time \cite{OSIappnote} is given by:

\begin{equation}
    \label{eqn:diff}
    \tau_{diff}= \frac{q(W_s-W_d)^2}{\mu_h k T}
\end{equation}

where $\mu_h$ is the hole mobility equal to 480 cm\textsuperscript{2}(Vs)\textsuperscript{-1} \cite{Knoll}.

A summary of the photodiode model parameters are given in Table \ref{tab:siparams}.

The PiN photodiode response curve is given by
\begin{equation}
    C(t)=-\exp{[-t/\tau_{fp}]}+\exp{[-t/\tau_{rp}]}
\end{equation}
where the photodiode fall-time $ \tau_{fp} > \tau_{rp} $.

The rise time governs the rate at which the photodiode can respond to a pulse of light. Typical literature values for the rise time are reported for the case where the photodiode is operated in its fully depleted mode at a reverse bias of 75 V. {It was therefore necessary to obtain a rise time for the conditions employed in practice. This was completed by setting the reverse bias to 12 V and using Equations \ref{photorise}-\ref{eqn:diff} with the above calculated $W_s$ and $N_n$}. A rise time of 15.08 $\mu$s was obtained as a result.  
The photodiode fall time was determined by matching the ADC pulse height for the \textsuperscript{137}Cs 662 keV energy peak in the simulation and the calibration and remained fixed at this value. This parameter, ultimately, serves as a ``catch-all'' for inaccuracies which might arise in the estimation of the doping concentration of silicon, or the substrate width. Therefore, while model parameters have been chosen in Table \ref{tab:siparams}, there are other possible combinations which yield the same rise time. 

\begin{table}
    \caption{Photodiode Model Parameters}
    \centering
    \begin{tabular}{|l|c|c|c|}
    \hline
        Parameter & Symbol & Value & Source \\
        \hline 
         Silicon doping concentration & $N_n$ & 2.985 $\times$ 10\textsuperscript{18}  & Estimated from \cite{osipin}\\
        Substrate width & $W_s$ & 210.7252 $\mu$m & Estimated from \cite{osipin}\\
        Load resistance & $R_L$ & 50  $\Omega$ & Estimated \cite{osipin} \\
        Contact resistance & $R_c$ & 0 $\Omega$ & Assumed\\
        Electron mobility & $\mu$ & 1350 $\mathrm{cm^{2}(Vs)^{-1}}$ & Estimated \cite{Knoll} \\
        Hole mobility & $\mu_h$ & 480 $\mathrm{cm^{2}(Vs)^{-1}}$ & Estimated \cite{Knoll} \\
        Silicon built-in bias & {$V_{Bi}$} & 0.65 V & Estimated\\
        Photodiode rise time & $\tau_{rp}$ & 15.08 $\mu$s & Derived\\
        Photodiode fall time & $\tau_{fp}$ & 32 $\mu$s & Empirical\\
        \hline
    \end{tabular}
    \label{tab:siparams}
\end{table}

\subsection{Current Pulse}
\label{cre}
The current pulse is given by a convolution of the scintillator and photodiode response, using the photodiode responsivity - digitised and shown in Figure \ref{fig:emissrespon} - to convert from the scintillator power to electrical current \cite{OSI}. The current pulse, $I(t)$ is given by:

\begin{equation}
    I(t)=R_{\lambda}P(t)*C(t)
\end{equation}
where $R_{\lambda}$ is the photodiode responsivity over the same wavelength range as the scintillator emission spectrum. The responsivity is the ratio of current generated to optical power. The magnitude of the current pulse is given explicitly by:

\begin{equation}
    I_{max}=\frac{1}{|L(t)*C(t)|}\sum_{\lambda=400}^{800} \frac{E_s(\lambda)N(\lambda)R_{\lambda}(\lambda)}{t_{co}}
\end{equation}

Where, $|L(t)*C(t)|$, denotes the maximum value of the convolution therefore serving as a normalisation factor. 

\subsection{Electronics Simulation}
\label{ltsec}
 The current pulse from the physics model has been used as a current source input in LTspice. The current source is connected to the transimpedence amplifier, yielding a voltage trace. The remainder of the circuit - until the microcontroller - was built in LTspice, following the layout given in Figure \ref{fig:blockdiag}. The measurement of the pulse analogue and valley detector voltages were completed manually, converting the voltage into an analogue-to-digital count. By varying the input energy, a relationship between incident energy and pulse height was determined. 

\section{Laboratory Calibration}
\label{labcal}
\begin{figure}
    \centering
    \begin{subfigure}{0.48\textwidth}
        \includegraphics[width=\textwidth]{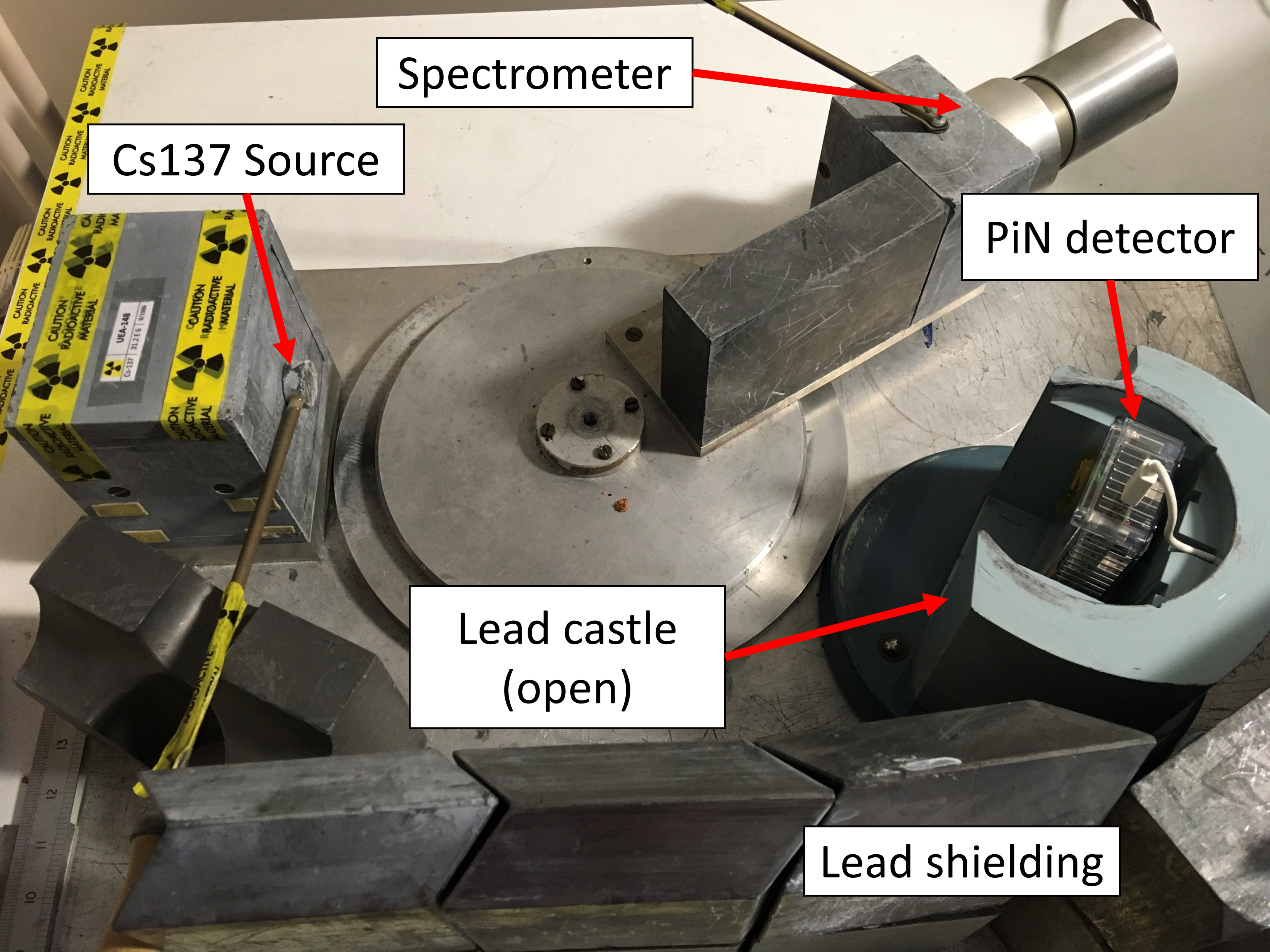}
        \caption{Photograph of the experiment. To collect \textsuperscript{137}Cs data, {each instrument was independently placed such they were directly opposite the source, as shown in Figure \ref{fig:Bathsetup}.}}
        \label{fig:bathmain}
    \end{subfigure}
    \hfill
    \centering
    \begin{subfigure}{0.48\textwidth}
        \includegraphics[width=\textwidth]{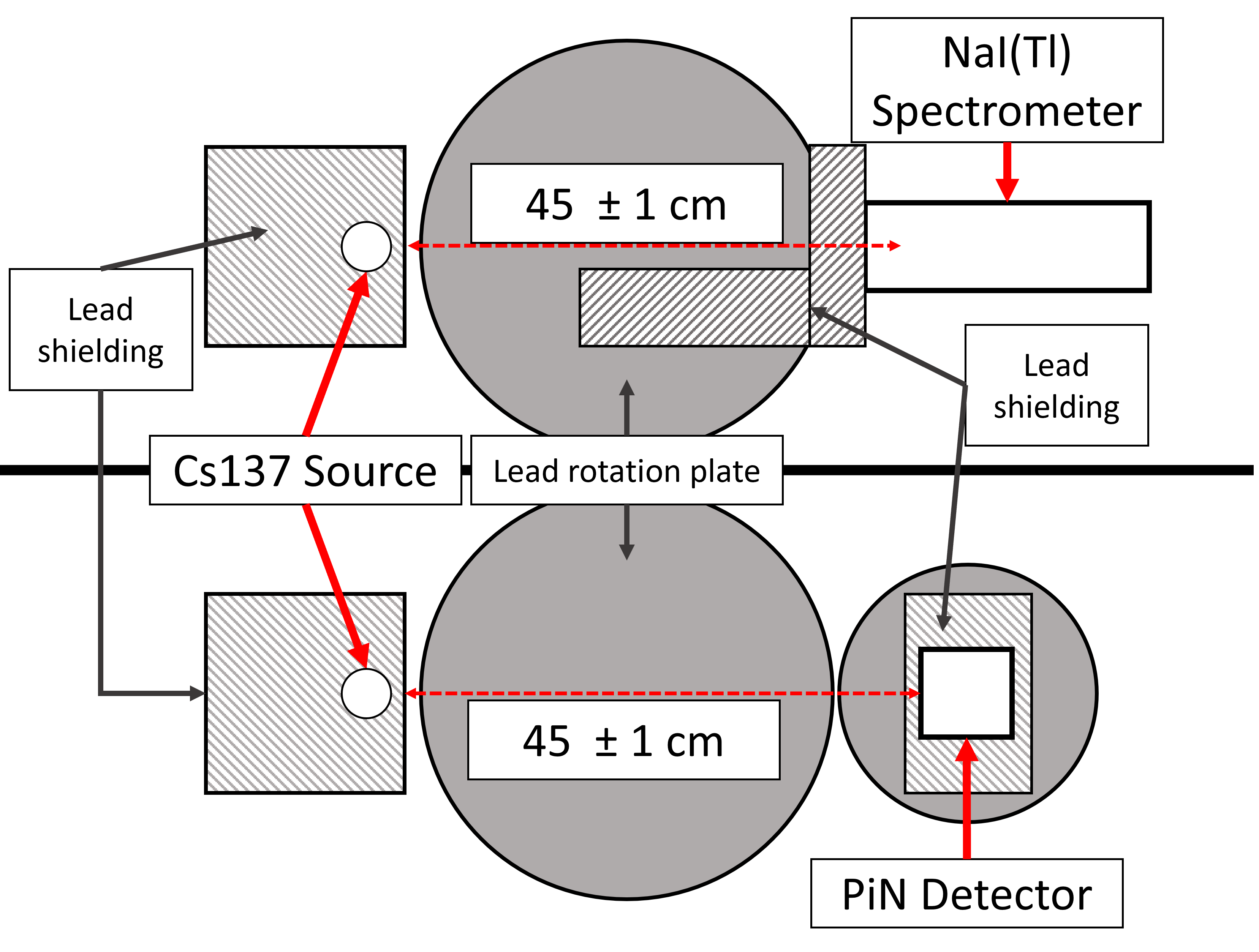}
        \caption{{Experimental setup during data collection. The top shows the configuration when the spectrometer was in use, and the bottom shows the configuration for the PiN detector. In the case of the PiN detector the top of the lead enclosure was in place, whereas in Figure \ref{fig:bathmain} it was removed.}}
        \label{fig:Bathsetup}
    \end{subfigure}
    \caption{{Experimental setups for the collection of \textsuperscript{137}Cs spectra data.}}
    \label{BathLab}
\end{figure}

To validate the model, {a PiN detector was calibrated using a Cesium-137 (\textsuperscript{137}Cs) radioactive source, and a P-2000 NaI(Tl) spectrometer. The P-2000 uses a 3.8 cm diameter by 2.5 cm thick NaI(Tl) crystal coupled to a photomultiplier tube (PMT) with diameter of 5.1 cm. The PMT is connected to a 1 kV multi-channel analyser, with data recorded via the Genie2000 software. The probe reports a resolution of 8.5 \%. The experimental setup for the \textsuperscript{137}Cs source can be seen in Figure \ref{fig:bathmain}}. 
{During \textsuperscript{137}Cs data collection, both instruments were aligned as seen in Figure \ref{fig:Bathsetup}. The activity of the \textsuperscript{137}Cs source was $15\times10^6$ Bq and both instruments were placed 45 $\pm$ 1 cm from the source. The PiN detector was exposed to the source for 287.96 seconds, and the spectrometer was exposed for 421 seconds.}

\begin{figure}[h!]
    \centering
    \includegraphics[width=\textwidth]{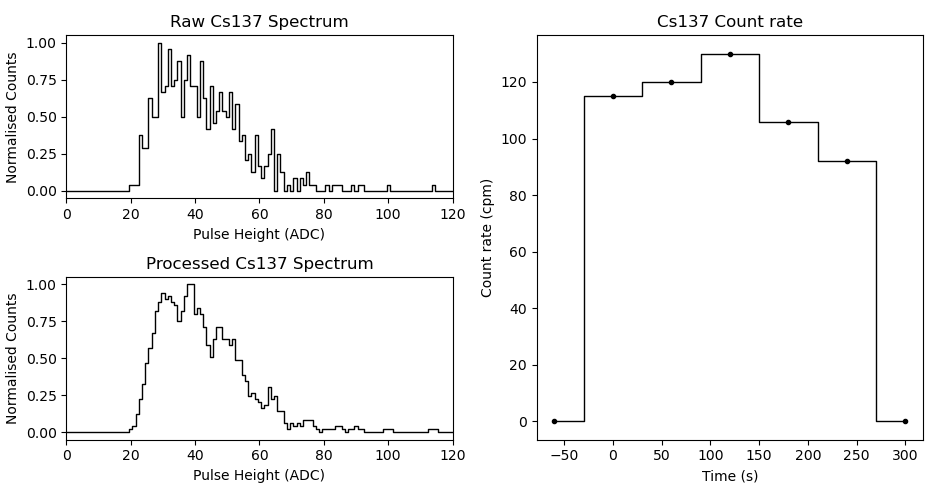}
    \caption{{Experimental data from the \textsuperscript{137}Cs source. In the raw spectrum, the pulse heights are sorted into 1-ADC-sized bins. The processed pulse height spectrum displays the average frequency when the entire distribution is shifted up, and down, 1 ADC. Through this process, the error on a pulse height measurement is considered, and energy peaks begin to emerge. The maximum recorded pulse height was 137 ADC. Additionally, the total counts per minute are provided. The average count rate over the exposure time was 117.1 cpm.}}
    \label{fig:PiNData}
\end{figure}

{Figure \ref{fig:PiNData} displays the raw binned PiN data from the \textsuperscript{137}Cs source, the processed spectrum, and the count rate from the data set. To obtain the processed spectrum, the error on a single pulse height ($\pm$1 ADC) is considered. From the list of pulse heights, a new list is created where each event is 1 ADC greater than its original value and another list where each event is 1 ADC smaller. These two lists and the original list are binned normally, and the frequency in each bin is averaged.} 

\begin{figure}
    \centering
    \includegraphics[width=\textwidth]{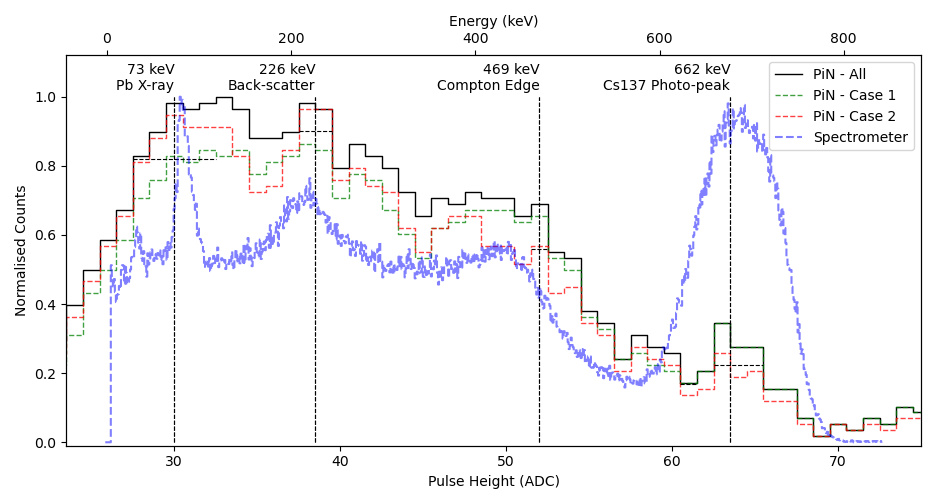}
    \caption{Summary of the detector calibration with a \textsuperscript{137}Cs source. The \textsuperscript{137}Cs energy spectrum measured with the P-2000 spectrometer is shown as a blue dashed line, with the corresponding energy axis at the top of the plot. The pulse height spectrum for all the PiN counts has been overlaid, with the corresponding axis shown at the bottom of the plot in analogue to digital converter (ADC) counts. Energy peaks have been identified, and their locations marked in the PiN pulse height domain, not the spectrometer energy domain. Case 1 denotes a subset of all PiN data, where the larger of two simultaneous events are considered as well as unique time events. Case 2 denotes the subset where the smaller of two simultaneous events are considered in addition to unique events.}
    \label{fig:PiNSpec}
\end{figure}

{\subsection{PiN Detector and P-2000 Spectrometer}}
{The P-2000 spectrometer allows for energy peaks in the PiN detector spectrum to be identified. {The spectrometer was first calibrated using the energy peaks detailed in Table \ref{tab:calipeaks} (R\textsuperscript{2} = 0.9909, p-value $\leq$ 0.0004); the errors on P-2000 energy values were given by the spectrometer calibration. Figure \ref{fig:PiNSpec} overlays the energy and pulse height spectra from each instrument. The two spectra were aligned by adjusting the pulse height limits in the PiN spectrum until its shape matched that of the NaI(Tl) spectrum.}
{The location of the features seen in the PiN spectrum were then compared to the energy at that location in the spectrometer. In Figure \ref{fig:PiNSpec}, due to the spectrometer calibration, the Compton edge of the spectrometer is at $\sim$430 keV; the marked vertical line at 469 keV is the energy at that location in the plot, where an edge can be seen in the PiN spectrum.}
\begin{table}[]
    \caption{{Characteristic energy peaks used for the calibration of the P-2000 spectrometer.}}
    \centering
    \begin{tabular}{|l|c|c|}
        \hline
        Origin & Characteristic Energy (keV) & P-2000 Energy (keV)\\
        \hline
        K-$\alpha$ X-ray (\textsuperscript{137}Ba) & 31 &  30 $\pm$ 20 \\
        Pb X-rays & 88 &  80 $\pm$ 30 \\
        Back-scatter Peak & 200 &  220 $\pm$ 30 \\
        Compton Edge & 477 & 430 $\pm$ 50 \\
        \textsuperscript{137}Cs Photo-peak  & 662 & 680 $\pm$ 60 \\
        \hline
    \end{tabular}
    \label{tab:calipeaks}
\end{table}

Additionally, Figure \ref{fig:PiNSpec} contains two subsets of data referred to as case 1 and case 2. In the \textsuperscript{137}Cs data, 60 pairs of events (120/562 events total) occur at the same timestamp, i.e. are simultaneous to less than 100 $\mu$s. There are several possible reasons: detector error through double-counting, coincidence, or a gamma and beta, which can be emitted simultaneously for some events in small detectors \cite{Knoll}. To deal with this two approaches have been taken to this data: case 1 displays all the `unique' time events and the larger event of each of the 60 pairs (i.e. it ignores the smaller of the `simultaneous' detections). Case 2 displays the unique time events and the smaller of each pair. While beyond the scope of this study, it is interesting to note how well case 2 matches the low energy region of the spectrometer spectrum, suggesting it could be due to the small detector effect of scattered beta particles.}

{An energy calibration of the PiN detector was completed by linear regression using the energy peaks identified in Figure \ref{fig:PiNSpec}}.

{The energy calibration, giving the relationship between pulse height ($PH$ in ADC) and output energy, is given below where $E_{LR}$ (in keV) is energy determined via linear regression:} 
\begin{equation}
    {E_{LR}=(17.6\pm0.2)PH-(450\pm10)}
\end{equation}
{with R\textsuperscript{2} = 0.9997 and p-value $\leq$ 0.0001.}

\section{Comparison of Detector and Model}
\label{comp}
\subsection{Simulation and Laboratory Results}
The {laboratory} calibration allows for pulses caused by specific gamma energies to be identified. {A further verification of the simulation can be obtained by using the gamma peaks identified in the lab as model inputs to predict the detector's peaks. The relationship between the model input energy $E_{Sim}$ (keV) and the pulse height  (ADC) is given by:} 

\begin{equation}
    {{E_{Sim}=(17\pm2)PH-(420\pm80)}}
\end{equation}

\begin{figure}
    \centering
    \includegraphics[width=\textwidth]{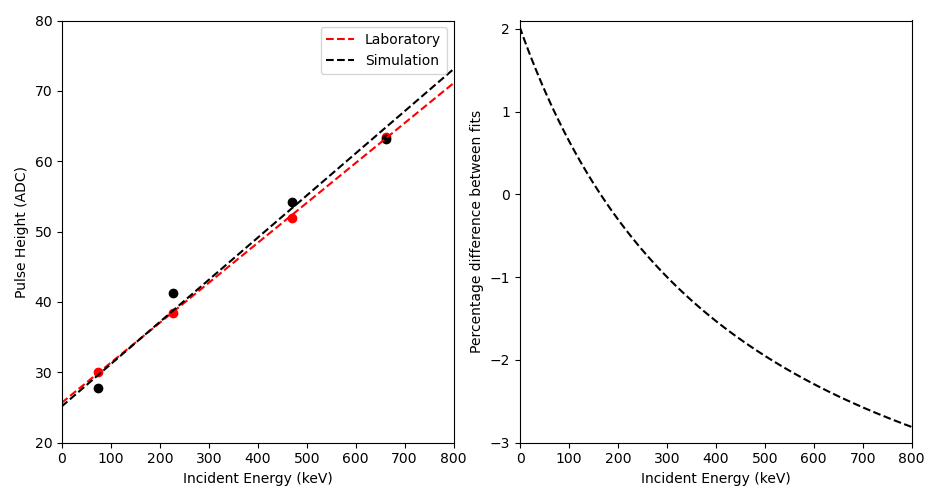}
    \caption{{Left:} Energy calibrations for the simulated and laboratory data sets. For the simulation: R\textsuperscript{2}-value = 0.9823, p-value $\leq$ 0.009; for the laboratory calibration: R\textsuperscript{2}-value = 0.9997, p-value $\leq$ 0.0001.
    {Right:} Percentage difference between the calibration and simulations linear fits from {0 - 800} keV input energy; over this energy range, the percentage difference {spans 2.00 to -2.81} \%.}
    \label{fig:simcal}
\end{figure}

{This linear fit has an R\textsuperscript{2} = {$0.9823$ and p-value $\leq$ $0.009$}. Figure \ref{fig:simcal} shows the selected energy peaks and the pulse heights as measured and predicted. In the 0-800 keV energy range, the percentage difference between measurement and prediction is between 2.00 \% and -2.81 \%.} 

The laboratory calibration yields an energy resolution of {17 $\pm$ 2} keV/ADC and sensitivity of {0.29 $\pm$ 0.03} mV/keV.

\subsection{Detectable Energy Range}
{The smallest detectable pulse height corresponds to the minimum detectable energy. There are several ways to determine this. A physics based approach would be to consider the energies of particles that can pass through the screening can in which the assembly is housed. The enclosure allows through 95 \% of 160 keV gamma rays, but 160 keV ($\sim$ 34 ADC) is much higher than the minimum recorded pulse height, therefore this threshold is not considered to be the limiting factor.\\ 
An electronics based approach would be to consider how pulses are measured. When the Schmitt trigger activates (dropping from 5 V to 0 V as shown in Figure \ref{fig:blockdiag}), a microcontroller interrupt starts the measurement of the valley and reference voltages. For some energies, the Schmitt trigger partially activates, dropping from 5 V to a voltage between 5 and 0 V. For the Schmitt trigger to activate the measurement interrupt, the input pulse must reach the trigger threshold, determined by the circuitry \cite{Horowitz2015}. Therefore, the minimum energy corresponds to the smallest pulse that will fully activate the trigger.} 

\begin{figure}
    \centering
    \includegraphics[width=0.75\textwidth]{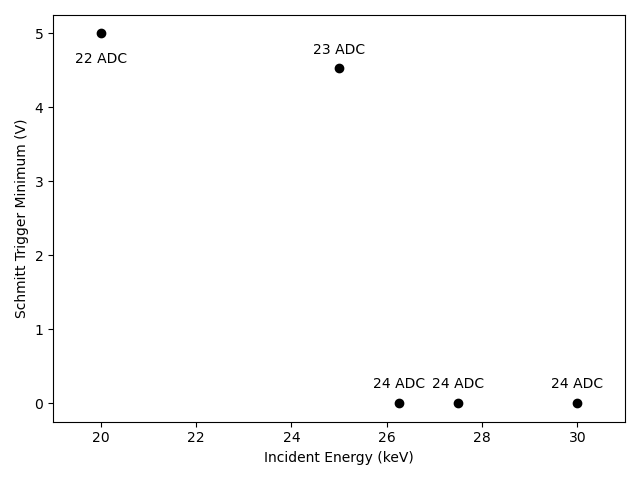}
    \caption{Detection threshold for low energy particles, determined from the minimum voltage reached by the Schmitt trigger in the LTspice simulation against input energy. For each energy, the calculated pulse height is shown in ADC counts. 0V would mean the trigger is activated, which occurs at {$\sim$26} keV, and the event will be registered by the microcontroller.
    }
    \label{fig:schmitt}
\end{figure}

 Figure \ref{fig:schmitt} shows the minimum value of the simulated Schmitt trigger for input energies between {20-30} keV. Over this range, the {two} lowest energies in Figure \ref{fig:schmitt} would not be detected because the Schmitt trigger has not activated.
 {The trigger is set to respond to pulses $\sim$20 ADC below the reference level to manage signal-to-noise. If the trigger were made more sensitive it would not be possible to distinguish between electrical noise and very low energy particles.}
 
 By using the model described in this paper, it can be seen that pulses corresponding to energy greater than or equal to {26} keV would be detected in the physical system, as there would be full activation of the trigger. Gamma rays with energy in the {24-26} keV range only partially activate the Schmitt trigger and are not detected. 

{In the \textsuperscript{137}Cs calibration experiment, the smallest pulse measured was 20 ADC, whereas Figure \ref{fig:schmitt} shows the smallest detectable pulse height in the simulation is 24 ADC. This difference between simulation and measurement is almost certainly caused by fluctuations in the reference voltage, which varies by typically $\pm$ 2 ADC. The detection threshold of 20 ADC corresponds to unphysical negative energies in the calibration experiments, which is most likely due to a lack of high resolution, low energy ($<$100 keV) peaks to optimise the calibration in this region.}

The maximum detectable energy could be governed by several aspects. Assuming normal behaviour of the reference level, the maximum pulse height could be 657 ADC, corresponding to an energy of {10-11} MeV, purely considering the available voltage drop. However, more likely the limiting factor on the maximum detectable energy would be the absorption efficiency of the scintillator. For similar sized CsI(Tl) crystals, the absorption efficiency has been shown to decrease to $\sim$ 10\% at 10 MeV \cite{FirstSen}. The maximum detectable energy is therefore assumed to be $\sim$ 10 MeV.

In summary, the trigger threshold voltage currently determines the minimum detectable energy. The threshold value could be adjusted to increase sensitivity to lower-energy particles, but this would need to be carefully traded off against the level of noise observed in the detector. The maximum detectable energy is controlled by the decrease in sensitivity of the detector at high energies, with the full range approximately 26 keV - 10 MeV. 

\subsection{Model considerations}
\label{modcon}
Due to the commercial nature of the photodiode used in the detector, a number of the parameters had to estimated. Some combinations of parameters result in the same predicted response. For example, if the scintillator cutoff time is decreased, producing a larger optical power, or the photodiode fall time increased, or the photodiode rise time decreased, all of these increase the size of the final pulse for constant input energy. The photodiode rise time in turn is dependent on the doping concentration, electron and hole mobilities, and resistances in the diode circuit. Therefore, flexibility is available in these model parameters to fine tune its performance to more closely match the actual system. 

\begin{figure}
    \centering
    \includegraphics[width=0.75\textwidth]{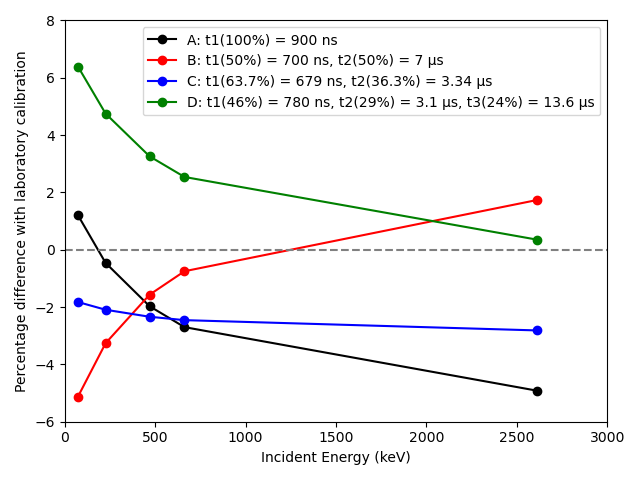}
    \caption{{A-D denote the different sets of decay times. For each set, the decay time(s) and proportion(s) are provided. Set A is the model response with the Advatech decay time, and demonstrates a -5.0 to 1.3 $\%$ deviation from the laboratory calibration. Set B spans -5.2 to 1.8 \%. Set C spans -2.9 to -1.8 \%. Set D, which is composed of three decay times, demonstrates an upper percentage difference of 6.4 \%, and a lower of 0.3 \%.} The final data point for each set of parameters shows the response to a 2614 keV gamma ray.}
    \label{fig:ddecay}
\end{figure}

{Additionally, CsI(Tl) crystals are often described by different decay times. We have considered the model sensitivity to three sets of decay parameters. Figure \ref{fig:ddecay} shows the difference between the laboratory calibration, and calibrations obtained from the three sets of additional decay times (B\cite{CRANNELL1974253}, C\cite{Valentine} \& D\cite{setD}), and the initial Advatech single decay time (A).}

{For set C, the absolute percentage difference over this energy range is $\sim$1.1 \%, suggesting that the scintillator in the PiN detector could be more accurately described by these decay times than a single decay time. However, rather than pick an arbitrary set of decays, the single decay time is retained in the model. Set B spans the largest range of deviation ($\sim$7 \%), however in the mid energy range (500-2000 keV), these parameters lie closest to the laboratory calibration.} The large deviation of set D at the low energy range suggests that this model should not be used for scintillator crystals described by similar decay times. Further work could more accurately determine the decay time constants for the scintillator sensor used.

\section{Conclusion}
A model has been developed to emulate the sensor and electronics response of a miniaturised CsI(Tl) PiN photodiode ionisation detector.  
The rise and decay times for the scintillator are specified as 50 ns and 900 ns, respectively.
The final photodiode rise time was determined to be 15.08 $\mu$s and fall time was determined to be {35} \textrm{$\mu$}s. 
The combination of a physics model and an electronics simulation have resulted in a full detector model which has been verified against {a} laboratory calibration with a \textsuperscript{137}Cs source. At {100} keV the difference between the model and laboratory data was {0.64} \%, and at 800 keV, it was {-2.81} \%. This modelling plus new calibration techniques have led to improvements in the detector energy resolution compared to previous work \cite{KA2017}\cite{KA2021}, to typically {17} $\pm$ {1} keV, in a range of {100} - 800 keV. This is suitable for the intended meteorological radiosonde application, as data rate limitations typically further degrade the energy resolution.

The model has been used to suggest that the Schmitt trigger threshold voltage appears to determine the minimum detectable energy of approximately 26 keV.
Agreement between the predicted and measured response means that components in the electronics can confidently be changed and tested in the simulation before implementation, speeding up development. 
Despite the PiN photodiode being a commercial product, with some physical quantities kept as proprietary information, the model affords flexibility in choosing, and matching, the photodiode fall time to accommodate parameters which have otherwise been estimated or assumptions which may not hold. It can also be adapted to simulate changes in the system such as different scintillator material or the response at different temperatures.

Overall this investigation offers increased confidence in the response of the detector, especially when considering the minimum detectable energy. This aspect is particularly valuable when evaluating the atmospheric effects of bremstrahhlung X-rays from energetic electron precipitation during space weather events \cite{KA2021}. Overall the model and laboratory calibration were consistent to {an absolute difference} better than {5}\% over the {calibration} range (0-800 keV) of the instrument. The detailed understanding now acquired of this novel miniaturised instrument will allow for more effective analysis and interpretation of existing data and future design improvements.
\\
\\
\textbf{Data availability}

Data is available at Tabbett, Justin; Aplin, Karen (2023), “PiN Modelling Data”, Mendeley Data, V1, doi: 10.17632/mctr9ksk4r.1\\
\textbf{Acknowledgements}

We would like to thank Dr Alessandro Narduzzo at the Department of Physics, University of Bath for access to radioactive sources.\\
\textbf{Funding}

EPSRC studentship and A-Squared Technologies Ltd.

\bibliographystyle{elsarticle-num-names} 
\bibliography{References}

\end{document}